\documentclass[conference]{IEEEtran}
\IEEEoverridecommandlockouts

\usepackage{cite}
\usepackage{amsmath,amssymb,amsfonts}
\usepackage{algorithmic}
\usepackage{graphicx}
\usepackage{textcomp}
\usepackage[dvipsnames]{xcolor}
\usepackage{acro}
\usepackage{caption}
\usepackage{hyperref}
\usepackage{tikz}
\usepackage{pgfplots}
\pgfplotsset{compat=1.18}
\usepackage[caption=false]{subfig}
\usepackage[normalem]{ulem}
\usepackage{rotating}
\usepackage{tabularray}
\usepackage{microtype}

\usetikzlibrary{positioning,calc,arrows.meta,shapes.arrows,petri} 
\tikzset{
      pre/.style={<-, >=Triangle},
      post/.style={->, >=Triangle},
      inhibitor/.style={<-, >=Circle},
      reset/.style={<-, >={Triangle}{Triangle}},
    }

\usepackage{float}
\newfloat{listing}{tbp}{lop}
\floatname{listing}{Lst.}
\floatstyle{plaintop}
\restylefloat{listing}

\usepackage{listings}

\def\BibTeX{{\rm B\kern-.05em{\sc i\kern-.025em b}\kern-.08em
    T\kern-.1667em\lower.7ex\hbox{E}\kern-.125emX}}

\DeclareAcronym{dsl}{
  short = DSL,
  long  = domain-specific language
}

\DeclareAcronym{sva}{
  short = SVA,
  long  = SystemVerilog Assertions
}

\DeclareAcronym{hpc}{
  short = HPC,
  long  = high-performance computing
}

\DeclareAcronym{fsm}{
  short = FSM,
  long  = finite state machine
}

\DeclareAcronym{dut}{
  short = DUT,
  long  = device under test
}

\DeclareAcronym{uvm}{
  short = UVM,
  long  = Universal Verification Methodology
}

\DeclareAcronym{dram}{
  short = DRAM,
  long  = dynamic random access memory
}

\DeclareAcronym{jedec}{
  short = JEDEC,
  long  = Joint Electron Devices Engineering Councils
}

\DeclareAcronym{hbm}{
  short = HBM,
  long  = high-bandwidth memory
}

\DeclareAcronym{ddr}{
  short = DDR,
  long  = double data rate
}

\DeclareAcronym{lpddr}{
  short = LPDDR,
  long  = low-power double data rate
}

\DeclareAcronym{rtl}{
  short = RTL,
  long  = register-transfer level
}

\newcommand{\myarctj}{
        \begin{tikzpicture}[baseline={([yshift=-1.5ex]current bounding box.center)}]    
            \draw[-{Diamond}] (0,0) --  node [above]{\footnotesize$t_x$} (0.7,0);
        \end{tikzpicture}
}

\begin{document}

\title{DRAMPyML: A Formal Description of DRAM Protocols with Timed Petri Nets
 \thanks{This work was partly funded by the Federal Ministry of Research, Technology and Space (BMFTR) under grants 16ME0936 and 16ME0934K (DI-DERAMSys).}
}

\makeatletter 
\newcommand{\linebreakand}{%
    \end{@IEEEauthorhalign}
    \hfill\mbox{}\par
    \mbox{}\hfill\begin{@IEEEauthorhalign}
}
\makeatother 

\author{
\IEEEauthorblockN{Derek Christ}
\IEEEauthorblockA{
    \textit{JMU Würzburg}\\
    Würzburg, Germany \\
    derek.christ@uni-wuerzburg.de
}
\and
\IEEEauthorblockN{Thomas Zimmermann}
\IEEEauthorblockA{
    \textit{Fraunhofer IESE}\\
    Kaiserslautern, Germany \\
    thomas.zimmermann@iese.fraunhofer.de
}
\and
\IEEEauthorblockN{Philippe Barbie}
\IEEEauthorblockA{
    \textit{Fraunhofer IESE}\\
    Kaiserslautern, Germany \\
    philippe.barbie@iese.fraunhofer.de
}
\linebreakand
\IEEEauthorblockN{Dmitri Saberi}
\IEEEauthorblockA{
    \textit{Normal Computing}\\
    New York, NY \\
    dmitri@normalcomputing.ai
}
\and
\IEEEauthorblockN{Yao Yin}
\IEEEauthorblockA{
    \textit{Normal Computing}\\
    New York, NY \\
    yao@normalcomputing.ai
}
\and
\IEEEauthorblockN{Matthias Jung}
\IEEEauthorblockA{
    \textit{JMU Würzburg, Fraunhofer IESE}\\
    Würzburg, Germany \\
    m.jung@uni-wuerzburg.de
}
}

\maketitle

\begin{abstract}
The JEDEC committee defines various domain-specific DRAM standards.
These standards feature increasingly complex protocol specifications, which are detailed in numerous timing diagrams and command tables.
As new features and complex device hierarchies emerge, understanding these protocols becomes increasingly difficult without an expressive model.
While each JEDEC standard features a simplified state machine, it fails to reflect the parallel operation of memory banks and is in itself insufficient for describing the complex DRAM protocol.

In this paper, we present an evolved modeling approach based on timed Petri nets and Python as a programming language.
This model provides a more accurate representation of DRAM protocols, making them easier to understand and directly executable, which enables the evaluation of interesting metrics and the verification of controller RTL models, DRAM logic and memory simulators.
\end{abstract}


\vspace{-10pt}

\section{Introduction}
In recent years, there has been a significant increase in memory-intensive applications, especially in fields such as machine learning and \ac{hpc}.
These applications demand higher memory bandwidth, which places a greater focus on the memory subsystem.
To address these demands, \ac{jedec} has introduced a variety of domain-specific \ac{dram} standards, as illustrated in Figure~\ref{fig:jedec_standards}.
These standards include \ac{ddr}, \ac{lpddr}, and \ac{hbm}, which are tailored to meet the performance and efficiency needs of different application scenarios.
However, adopting these modern standards introduces significant challenges.
Memory controller IP designers must adhere strictly to the increasingly complex and evolving protocol specifications.
For example, recent generations of \ac{dram} have introduced features such as SameBank commands (DDR5) and Per2Bank commands (LPDDR5), as well as multi-level organizational hierarchies, including stacks and pseudo-channels in the case of HBM, and logical and physical rank organizations in the case of DDR5.
\begin{figure}
    \centering
    \includegraphics[width=0.9\linewidth]{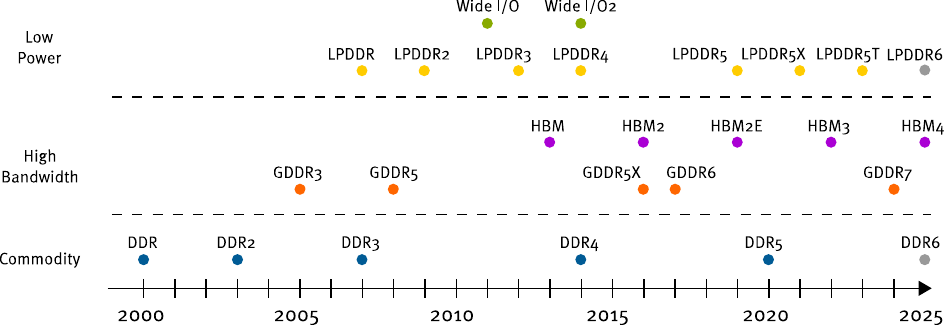}
    \caption{JEDEC standard releases over time~\cite{junkra_19}.}
    \label{fig:jedec_standards}
\end{figure}
These features significantly increase the difficulty of understanding, designing for, and verifying compliance with \ac{dram} protocols, as well as adhering to the complex timing dependencies between commands and component hierarchies.
Additionally, the standards released by \ac{jedec} are written without a formal structure.
This requires system architects to review timing diagrams, command tables, and the descriptions of all commands to understand how features are implemented and which commands depend on which others.
Consequently, if the system architect misinterprets or overlooks certain details of the standard, there is much room for error.
Several previous works have attempted to formalize complex \ac{dram} protocol descriptions, such as in~\cite{junkra_17, junkra_19, stesud_22}.
However, these efforts have failed to keep up with the growing complexity of newer standards because they use a rigid \ac{dsl} called \textit{DRAMml} that is unable to express SameBank or Per2Bank commands.
Additionally, the strong hierarchies imposed by the custom language makes expressing the diverse component organizations in modern standards difficult.
However, this formal description provides a strong basis to build upon:
Together with Jetbrains~MPS~\cite{jet_25}, the authors used the language to generate source code for the DRAMSys~\cite{stejun_22a} cycle-accurate \ac{dram} simulator, resulting in correct-by-construction standard compliance.
Listing~\ref{lst:dramml} shows the DRAMml description of a $t_{RCD}$ timing constraint between an ACT and an RD command, which results in the generated code in Listing~\ref{lst:timing_constraint}.
In addition, the same description was used to automatically generate \ac{sva} code (see Listing~\ref{lst:sva}), which verifies whether existing \ac{rtl} memory controllers comply with the standard.
Therefore, it is evident that the DRAMml description has great potential for formalizing and analyzing future JEDEC standards.
\begin{listing}
\footnotesize
\begin{lstlisting}
TimingConstraints {
    Arcs {
        ACT -<> RD (tRCD,0);
        // ...
    }
}
\end{lstlisting}
\caption{DRAMml description of a $t_{RCD}$ timing constraint.}
\label{lst:dramml}
\end{listing}
\begin{listing}
\footnotesize
\begin{lstlisting}
sc_time c = sc_time_stamp() + tRCD;
sc_time &nextPossibleTime = nextCmdByBank[RD][bank];
nextPossibleTime = std::max(nextPossibleTime, c);
\end{lstlisting}
\caption{DRAMSys timing constraint for $t_{RCD}$.}
\label{lst:timing_constraint}
\end{listing}
\begin{listing}
\footnotesize
\begin{lstlisting}
property arc_active_act;
    @(posedge clk) disable iff (reset)
        (cmd == act) |-> (active >= 1'b1);
end property;
assert property(arc_active_act);
\end{lstlisting}
\caption{\acl{sva} property of a place-to-transition arc for $t_{RCD}$.}
\label{lst:sva}
\end{listing}
However, the technical foundation requires an update in order to express upcoming complex timing dependencies and device hierarchies with greater flexibility and ease of maintenance.
In this paper, we present a new formal description, \textit{DRAMPyML}, which is based on the general-purpose scripting language Python.
DRAMPyML uses timed Petri nets to express JEDEC DRAM protocols, capturing not only functional correctness, but also timing dependencies between commands.
In summary, we make the following contributions:
\begin{itemize}
    \item A new formal description method for DRAM protocols based on Python and timed Petri nets is introduced.
    \item The model is made executable, and a simple protocol-compliant memory controller is created.
    \item It is demonstrated how the executable model can be used for verification of memory controllers and DRAM logic.
    \item A set of formal analyses of properties is performed with the help of the Petri net model.
\end{itemize}
Our hope is to set a standard for expressing DRAM protocols using the proposed timed Petri nets.

The remaining paper is structured as follows:
Section~\ref{sec:related_work} gives an overview of related work.
Section~\ref{sec:background} provides the necessary background information.
In Section~\ref{sec:implementation} the implementation of new method and the algorithms for certain formal analyses are described.
Section~\ref{sec:results} presents experimental results of the new model, and in Section~\ref{sec:conclusion} the paper is concluded.

\section{Related Work}
\label{sec:related_work}

Petri nets have proven useful for modeling systems of varying degrees of complexity~\cite{kheahm_23,zurdil_91}.
Past research has also investigated the use of Petri nets for modeling DRAMs.
Gries~\cite{gri_00} used Petri nets to model the internal workings of a DRAM, focusing on concrete memory cells and building a complete, bottom-up model that included all aspects of a DRAM device.
However, this level of detail is excessive for modeling and verifying communication with the DRAM through the command protocol.
In~\cite{fanell_03}, Petri nets were used to model DRAM power-down states to derive effective power-down strategies; however, the work solely focuses on power-down.
The authors of~\cite{junkra_17} used Petri nets to provide a comprehensive, formal model of the DRAM's internal state and the JEDEC-defined command transitions.
Petri nets avoid a state explosion by using place tokens to compress combinations of states, unlike regular \acp{fsm}.
However, this model only considers legal state transitions.
It does not account for the complex timing dependencies between commands that the DRAM controller must follow.
In~\cite{junkra_19}, the authors extend the Petri net model introduced in~\cite{junkra_17} by including custom timing arcs between transitions, which allows them to adhere to complex DRAM protocols from a timing perspective:
For example, the timing arc of $t_{RCD}$ between the transition corresponding to a bank's ACT command and the transitions of that bank's CAS commands models the necessary delay between row and column commands within the same bank.
Even more abstract concepts, such as the command bus occupation, refresh and the so-called \textit{Four Activate Window} are modeled with these primitives.
A \ac{dsl} called \textit{DRAMml} is used to express the defined Petri net.
DRAMml describes all timing, state, and command information of JEDEC standards in a formal and comprehensible format.
This work is further expanded upon in~\cite{stesud_22}, where the authors used the same DRAMml description to generate \ac{sva} properties:
A set of assertions is generated using the semantics of Petri nets and the DRAMml standard to formally verify the correctness of an RTL DRAM controller implementation or to identify protocol violations.
However, as DRAM standards become more complex, such as with DDR5's SameBank commands and LPDDR5's Per2Bank commands, DRAMml's strict hierarchical specification reaches its limits.
Therefore, it is imperative that the description of the constructed Petri net becomes more flexible and powerful.
For this reason, we are introducing a new DRAM modeling implementation based on Python Petri nets to enable a quicker reaction to complex protocol changes.

\section{Background}
\label{sec:background}

In this section, we first introduce the basic structure and terminology of DRAM before giving an overview of timed Petri nets in the context of modeling DRAM devices.  

\subsection{DRAM Basics}

In general, DRAM is organized into a hierarchy of channels, ranks, and devices.
The devices contain banks with rows and columns.
More recent standards have introduced additional hierarchy levels, such as bank groups (DDR4), logical ranks (DDR5), and pseudo-channels (HBM2).
These levels of hierarchy come with different degrees of independence.
While channels are completely separate entities with independent command/address and data buses, each channel can consist of multiple ranks that share the same buses to the controller.
The same is true for the devices on those ranks. The banks of a device can be used concurrently (\textit{bank parallelism}), but their dependency on the same bus connection leads to bus-related timing constraints that must be respected.
 
\begin{table}
\centering
\caption{DRAM Commands}
\label{tab:commands}
\begin{tblr}{
  width = \linewidth,
  colspec = {X[1] X[2] X[10]},
  rowsep = 2pt,
  colsep = 3pt,
  cell{2}{1} = {r=2}{c},
  cell{2}{2} = {c},
  cell{3}{2} = {c},
  cell{4}{1} = {r=2}{c},
  cell{4}{2} = {c},
  cell{5}{2} = {c},
  cell{6}{1} = {r=4}{c},
  cell{6}{2} = {c},
  cell{7}{2} = {c},
  cell{8}{2} = {c},
  cell{9}{2} = {c},
  vlines,
  hline{1-2,4,6,10} = {-}{},
  hline{2,4,6} = {2}{-}{},
  hline{3,5,7-9} = {2-3}{},
}
\textbf{Type}                             & \textbf{Command}          & \textbf{Explanation}                                                                             \\
\begin{sideways}Row\end{sideways}         & \texttt{ACT}              & \textit{Activate}: A specific row in one bank is activated.                                      \\
                                          & \texttt{PRE}              & \textit{Precharge}: The currently activated row is closed and the bank is precharged.            \\
\begin{sideways}Column\end{sideways}      & \texttt{RD/RDA}           & \textit{Read/Read with Auto-Precharge}: Read from an activated row~(and precharge afterwards).   \\
                                          & \texttt{WR/WRA}           & \textit{Write/Write with Auto-Precharge}: Write to an activated row~(and precharge afterwards).  \\
\begin{sideways}Entire DRAM\end{sideways} & \texttt{PREA}             & \textit{Precharge All}: All active banks are precharged.                                         \\
                                          & \texttt{REFA}             & \textit{Auto-Refresh}: Refresh one or more rows in all banks.                                    \\
                                          & \texttt{PDE/PDX}          & \textit{Power-Down Entry/Exit}: Enters or exits the PDN mode.                                    \\
                                          & \texttt{SRE/SRX}          & \textit{Self-Refresh Entry/Exit}: Enters or exits the SREF mode.
\end{tblr}
\end{table}

\begin{table}
\centering
\caption{DRAM States}
\label{tab:states}
\begin{tblr}{
  width = \linewidth,
  colspec = {X[1] X[2] X[10]},
  rowsep = 2pt,
  colsep = 3pt,
  cell{2}{1} = {r=2}{c},
  cell{2}{2} = {c},
  cell{3}{2} = {c},
  cell{4}{1} = {r=2}{c},
  cell{4}{2} = {c},
  cell{5}{2} = {c},
  vlines,
  hlines,
}
\textbf{Type}                                  & \textbf{State}          & \textbf{Explanation}                                                                                                 \\
\begin{sideways}Normal\end{sideways}           & \textit{ACTIVE}         & At minimum one bank is active, no powerdown (\texttt{cke=1}), no internal refresh.                                   \\
                                               & \textit{IDLE}           & All banks are closed and precharged, no power-down (\texttt{cke=1}), no internal refresh.                            \\
\begin{sideways}Power-Saving\end{sideways}     & \textit{PDN}            & \textit{Power-Down}: Banks are active or precharged and no internal refresh (\texttt{cke=0}).                        \\
                                               & \textit{SREF}           & \textit{Self-Refresh}: All banks are precharged, the DRAM internal self-timed refresh is triggered (\texttt{cke=0}). 
\end{tblr}
\end{table}

When accessing the data, the rows and columns of a bank cannot be accessed directly.
Before any data can be read (\texttt{RD}) or written (\texttt{WR}), a row must be activated (\texttt{ACT}); however, this can only be done for one row per bank at a time.
Before another row can be activated, a precharge (\texttt{PRE}) must be performed.
Additionally, because DRAM is charge-based with leakage effects, memory cells must be refreshed (\texttt{REFA}) regularly to avoid data loss.
There are two power-saving modes: power-down (\texttt{PDN}), where refreshes must be issued by the controller, and self-refresh (\texttt{SREF}), where the DRAM refreshes itself.
Table~\ref{tab:commands} summarizes the most important DRAM commands.

The possible actions depend on the current state of the DRAM of which the most important ones are shown in Table~\ref{tab:states}.
For the state transitions, each JEDEC standard contains a \ac{fsm} showing the legal actions and state transitions.
However, even JEDEC admits, that this state diagram is not fully correct~\cite{jed_24}, as it does not account for bank parallelism, which would lead to a state explosion.
For this reason, timing constraints are also not covered there and must be extracted from timing diagrams and other sources in the respective standard.
Timing constraints may be related to minimum timing delays between two specific commands, the command bus occupancy, the \textit{N}-Activate Window (NAW), or related to refresh mechanisms~\cite{junkra_19}.

\subsection{Timed Petri Nets}
Petri nets~\cite{pet_62} allow for the modeling of concurrent, asynchronous systems using graphs consisting of \textit{places} that hold \textit{tokens}, \textit{transitions} that can be fired and \textit{arcs} which connect places to transitions or vice versa.
Typically, a place represents a condition or state while transitions represent events~\cite{mur_89}.
As an example, Figure~\ref{fig:petri_net_example} shows the Petri net model of a chemical reaction of $2H_2 + O_2 = 2H_2O$.
The tokens inside the places show the available units of each compartment, the weights at the arcs indicate the change to the number of units in the place when the transition is fired.

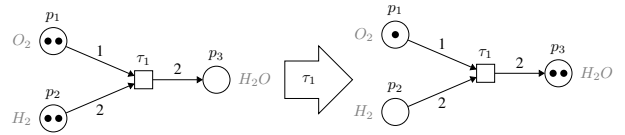
\begin{figure}
    \centering
    \scalebox{0.6}{
      \begin{tikzpicture}[node distance=1.1cm, place/.style={circle, draw, minimum size=6mm}]
    \node[place,tokens=2,label=above:$p_1$] (p1) {};
    \node[left=0.05cm of p1, text=gray] (o2) {$O_2$};
    \node[place,tokens=2,label=above:$p_2$,below=of p1] (p2) {};
    \node[left=0.05cm of p2, text=gray] (o2) {$H_2$};

    \node[transition,label=above:$\tau_1$] (t1) at ($(p1)!0.5!(p2) + (2,0)$) {}
    edge[pre] node[midway, above] {1} (p1)
    edge[pre] node[midway, below] {2} (p2);

    \node[place,label=above:$p_3$, right=of t1] (p3) {}
    edge[pre] node[midway, above] {2} (t1);
    \node[right=0.05cm of p3, text=gray] (o2) {$H_2O$};
    
    \node[single arrow, draw=black, label=center:$\tau_1$, minimum width=1.5cm, minimum height=1.5cm, single arrow head extend=1mm, right=of p3, xshift=0.1cm] (arr) {};
    \coordinate (help) at ($(arr)+(0.5,1)$);

    \node[place,tokens=1,label=above:$p_1$,right=of help] (p4) {};
    \node[left=0.05cm of p4, text=gray] (o2) {$O_2$};
    \node[place,tokens=0,label=above:$p_2$,below=of p4] (p5) {};
    \node[left=0.05cm of p5, text=gray] (o2) {$H_2$};
    
    \node[transition,label=above:$\tau_1$] (t2) at ($(p4)!0.5!(p5) + (2,0)$) {}
    edge[pre] node[midway, above] {1} (p4)
    edge[pre] node[midway, below] {2} (p5);

    \node[place,tokens=2,label=above:$p_3$, right=of t2] (p6) {}
    edge[pre] node[midway, above] {2} (t2);
    \node[right=0.05cm of p6, text=gray] (o2) {$H_2O$};
    
\end{tikzpicture}
    }
    \caption{Petri net example according to~\cite{mur_89}.}
    \label{fig:petri_net_example}
    
\end{figure}

To model DRAM devices, the authors of~\cite{junkra_17} have extended the Petri net to include \textit{Inhibitor-}~\cite{age_73,hack_76} and \textit{Reset Arcs}~\cite{ara_76}.
\textit{Reset Arcs} (\tikz{\draw[-{Triangle}{Triangle}, fill=black, draw=black, minimum size=1em, inner sep=0pt] (0,0) -- (0.5,0) node[ scale=0.5]{};}) are directed from a place to a transition and remove all tokens in the connected place when the transition is fired.
Meanwhile, an \textit{Inhibitor Arc} (\tikz{\draw[fill=black, draw=black, circle, minimum size=1em, inner sep=0pt] (0,0) -- (0.5,0) node[circle, fill, draw, inner sep=0pt, scale=0.5]{};}) is directed in the same way, but prevents the connected transition from firing when the number of tokens in the place is greater or equal to the weight of the arc\cite{verwyn_10}.
The model in Figure~\ref{fig:petri_net_bank} uses reset and inhibitor arcs to block \texttt{ACT} when \texttt{ACTIVE} has a token and to reset tokens when \texttt{PRE} fires.
As done in \cite{pet_81}, it can be proven that Petri nets with inhibitor arcs possess the modeling power of Turing machines.
Based on this, in ~\cite{junkra_17}, a DRAM Petri net model of a DRAM device is presented, where the transitions correspond to commands, places to events, and bank parallelism is managed through the use of tokens.

\begin{figure}
    \centering
    \includegraphics[width=0.4\linewidth]{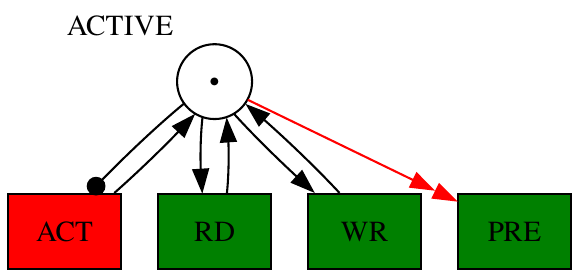}
    \caption{Petri net model of a single bank.}
    \label{fig:petri_net_bank}
\end{figure}

In~\cite{junkra_19}, this model was extended to include the timing dependencies listed in the previous section.
To this end, an additional extension was required, necessitating the use of \textit{Timed Arcs}~(\tikz{\draw[-{Triangle}, fill=black, draw=black, minimum size=1em, inner sep=0pt] (0,0) -- (1.3,0) node[midway, above, yshift=0.8]{\small $[t_1,t_2]$};})~\cite{jacjac_11,zub_91}.
The tokens were assigned an age, and the timed arcs were equipped with age guards.
This ensures that only tokens within a specified age range are taken into account along the arc and are capable of activating the associated transition.
Timed arcs enabled the modeling of the \textit{NAW} and the \textit{Command Bus Constraint}.
For the \textit{Command-To-Command Timing Dependencies} and the \textit{Refresh Mechanism}, the authors introduced a custom timing arc ($\myarctj$) based on a reset arc and a timed inhibitor arc:
In short, the arc indicates that firing the first transition inhibits the firing of the second transition for the duration of $t_x$.

To be able to adapt their \textit{Timed-Inhibitor-Reset Petri net} model to different standards, the authors of~\cite{junkra_19} developed a \ac{dsl} named \textit{DRAMml} based on Jetbrains MPS with a custom syntax:
A normal arc is denoted with \textbf{\texttt{->}}, while \mbox{\textbf{\texttt{-> [t1,t2]}}} denotes a timed arc with age guard $[t_1,t_2]$.
Reset arcs are denoted with \textbf{\texttt{->>}} and \textbf{\texttt{-o}} denotes an inhibitor arc.
The custom arcs denote using \mbox{\textbf{\texttt{-<> (t1,\dots,tn)}}} command-to-command constraints on different hierarchy levels.
 
The creation of a functional model can be accomplished through an automated process based on the \ac{dsl} description.
However, the strong hierarchy of this model renders it inflexible and difficult to adapt to newer standards, resulting in long and complex descriptions.
Consequently, a novel approach founded on Python is hereby presented for the formal description of the DRAM device as a Timed-Inhibitor-Reset Petri net, herein designated as \textit{DRAMPyML}.
This novel approach enhances flexibility and expandability while preserving the original approaches benefit of correctness-by-construction by generating the functional model from the description.

\section{Implementation}
\label{sec:implementation}
In this section, we present the implementation of the new flexible Petri net model.
First, the semantics are discussed in Section~\ref{sec:semantic}, and then the concrete syntax is shown in Section~\ref{sec:syntax}.
Finally, Section~\ref{sec:exec_model} describes algorithms used to perform a formal analysis of the model.

\subsection{Semantics}
\label{sec:semantic}
The general structure of the Petri net is based on the optimized topology used in~\cite{stesud_22}.
A simplified Petri net, loosely based on DDR3, with only one rank and two banks is depicted in Figure~\ref{fig:petri_net}.
\begin{figure*}
    \subfloat[Simplified Petri net with a single rank and two banks, with transitions that can fire (marked in \textcolor{OliveGreen}{green}) and that cannot fire (marked in \textcolor{Red}{red}).]{%
    \includegraphics[width=0.68\linewidth]{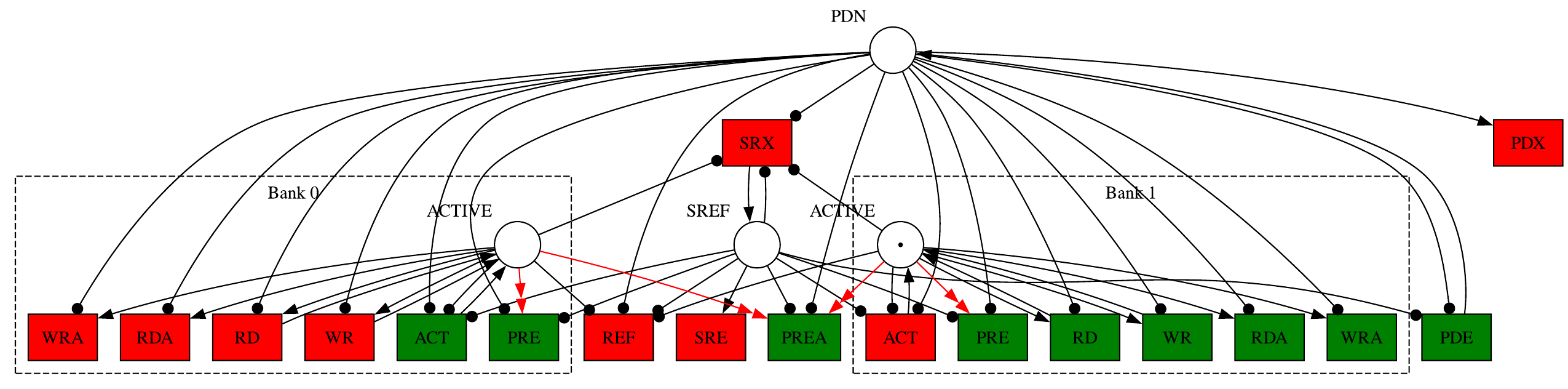}
    \label{fig:petri_net}
    }
    \hfill
    \subfloat[Reduced Petri net showing timing dependencies after firing \texttt{ACT} of Bank 0.]{%
    \includegraphics[width=0.28\linewidth]{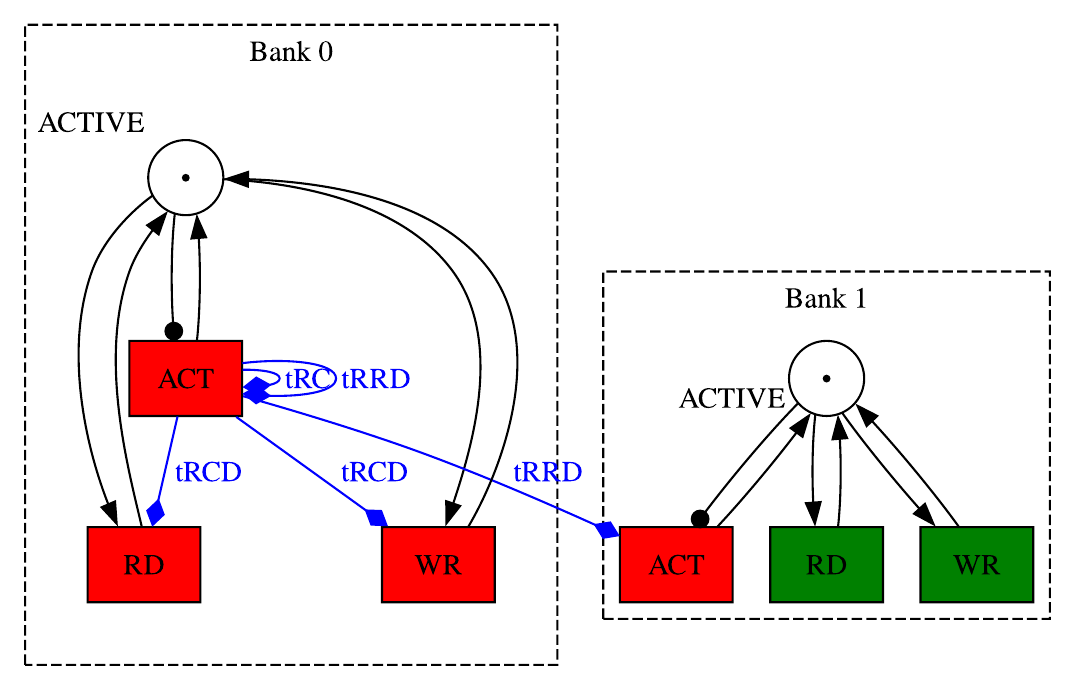}
    \label{fig:petri_net_timing}
    }
    \caption{Simplified DRAM Petri nets.}
\end{figure*}
In this model, there is an \texttt{ACTIVE} place for each bank that initially holds no tokens.
A token is only placed into the \texttt{ACTIVE} place when the respective \texttt{ACT} transition fires.
Note that firing \texttt{ACT} does not need to consume a token, but only generates one.
With the bank being active, its respective \texttt{RD}/\texttt{RDA} and \texttt{WR}/\texttt{WRA} transitions can be issued.
These transitions are getting inhibited by inhibitor arcs if the DRAM is in active power-down mode.
Similarly, the inhibitor arc from \texttt{ACTIVE} to \texttt{ACT} prevents the place from firing \texttt{ACT} again on the same bank without precharging first.
While one bank is active, the \texttt{REF} and \texttt{SRE} transitions are disabled because they can only be fired when all banks are precharged.
However, \texttt{PDE} can be issued to enter the the power-down mode with either all banks being precharged or one or more banks being active.
The power-down mode is defined here per rank; whether power-down can be controlled at the rank or channel level depends on the standard and if there are separate CKE signals for each rank.
Both the \texttt{PDX} and \texttt{SRX} transitions only consume a token from their respective place but do not generate a new one, leaving the rank in the idle state.
Due to the the reset arcs connected to the transitions, firing \texttt{PRE} or \texttt{PREA} deletes the token placed in \texttt{ACTIVE} and resets the corresponding banks to their idle state.

Using this model, all legal commands are expressed on a functional level.
However, the introduction of timing arcs between transitions makes it possible to adhere to even the most complex timing dependencies of the DRAM protocol.
In the following, the syntax of building of the described model is explained.
The new Python implementation makes use of the rustworkx~\cite{qis_25} graph library by Qiskit to build up the Petri net model in a performant, general-purpose graph data structure.
The rustworkx library provides Python bindings for the popular petgraph~\cite{pet_25} Rust crate, thus combining the performance of a compiled Rust data structure with the flexibility of Python for graph construction.

\subsection{Syntax}
\label{sec:syntax}

The nodes and edges of the rustworkx graph use custom data types to represent places, transitions, regular arcs, inhibitor arcs, reset arcs and timing arcs.
The definition of the general structure of the Petri net corresponding to Figure~\ref{fig:petri_net} is shown in Listing~\ref{lst:graph}.
\begin{listing}
    \footnotesize
    \caption{Definition of the general Petri net graph structure.}
    \begin{lstlisting}
g = rx.PyDiGraph()
for rank in range(numberOfRanks):
    # Add rank-wise places and transitions...
    for bank in range(numberOfBanks):
        # Add bank-wise places and transitions...
\end{lstlisting}
\label{lst:graph}
\end{listing}
A for loop iterates over the number of components at each level of the hierarchy (e.g., the number of banks) and initializes its places, arcs and transitions.
Listing~\ref{lst:nodes_edges} demonstrates the creation of places, transitions, and connections via arcs.
\begin{listing}
\caption{Exemplary creation of places, transitions and arcs in the Petri net.}
\footnotesize
\begin{lstlisting}
# ACTIVE place with an associated bank coordinate:
p_active = g.add_node(Place(ACTIVE, bank_coord))
# PREA transition with an associated rank coordinate:
t_prea = g.add_node(Transition(PREA, rank_coord))
# Reset arc from ACTIVE to PREA:
g.add_edge(p_active, t_prea, ResetArc())
\end{lstlisting}
\label{lst:nodes_edges}
\end{listing}
Since places and transitions correspond to ranks or banks, a coordinate is used to distinguish the \texttt{ACT} transition from one bank to another, for example.
A place is a special type of a node, and a transition is another type of a node, differentiated by its type.
In the data structure, arcs are represented as edges connecting nodes, i.e., they can connect places or transitions.

This model can also be used to represent timing constraints using timing arcs.
Listing~\ref{lst:pyml_timing_constraint} defines an exemplary timing constraint for $t_{RCD}$ at bank the level.
\begin{listing}
\footnotesize
\caption{Definition of the $t_{RCD}$ dependency for a bank.}
\begin{lstlisting}
CommandTimingConstraint(
    intra_bank, [ACT], [RD, WR, RDA, WRA], tRCD
)
\end{lstlisting}
\label{lst:pyml_timing_constraint}
\end{listing}
The written expression enables constraints to be defined for the Cartesian product $CMD_{from} \times CMD_{to}$ of two commands with the same timing value.
In this concrete example, $t_{RCD}$ applies between an \texttt{ACT} command and all column commands (\texttt{RD}, \texttt{RDA}, \texttt{WR} and \texttt{WRA}).
In a secondary step, the list of command constraints is iterated to instantiate the corresponding timing arcs between transitions in the graph.
The \texttt{intra\_bank} lambda function determines whether timing arcs are generated on the coordinates of the two transitions from the Cartesian product.
This flexible callback approach allows for the modeling of complex Per2Bank or SameBank constraints by evaluating them as \texttt{True} only if the constraint must be applied.

\subsection{Executable Model}
\label{sec:exec_model}

Implementing the Petri net with a general-purpose graph library enables the model to be made directly executable.
Figure~\ref{fig:petri_net_timing} illustrates a subsection of the Petri net, showing the timing dependencies after firing the bank's \texttt{ACT} transition to its CAS commands, as well as to the other bank's \texttt{ACT} transition.
However, the Petri net itself does not fire any transitions; it only defines the currently valid operations.
An external initiator (e.g., a memory controller or command trace) is required to trigger the firing of transitions.
With the help of special initiators, a set of algorithms can be applied to the graph to formally analyze the properties of the Petri net.
As each state is defined solely by the tokens in each place, it is possible to traverse the Petri net and generate all reachable states.
In other words, the Petri net can be unrolled into a \ac{fsm}.
However, due to state explosion, this is only feasible for simple nets with a small number of banks.
Nevertheless, the insights gained can also be applied to more complex bank organisations.
To implement such an algorithm, starting from the initial state, a list of all enabled transitions is generated.
After snapshotting the Petri net's current state, each of these transitions is then fired and the respective new state is compared to all previously reached states.
The transition itself is also recorded during this process to capture the edges between the states in the state diagram.

Another advantage of the executable model is that it allows all possible state transitions for a given depth $k$ of command sequences to be enumerated.
To achieve this, all fireable transitions originating from the idle state are explored in order to generate an initial path.
The same algorithm is then applied recursively with a depth of $k-1$.
This set contains all legal traces at a depth of $k$.
Therefore, any trace not contained within this set is not a valid command trace.
Such a set of valid command traces can be used to evaluate the equivalence of different Petri net topologies which may represent the same DRAM model.

\section{Results}
\label{sec:results}

In this section, we will apply the previously described algorithms to the simplified DRAM model and analyze the computational complexity.
Then, we demonstrate how the model can be used for verification and source code generation.

\subsection{Petri Net Unrolling}
Figure~\ref{fig:traversal} shows the unrolled Petri net for the previously analyzed example DRAM.
\begin{figure}
    \centering
    \includegraphics[width=0.8\linewidth]{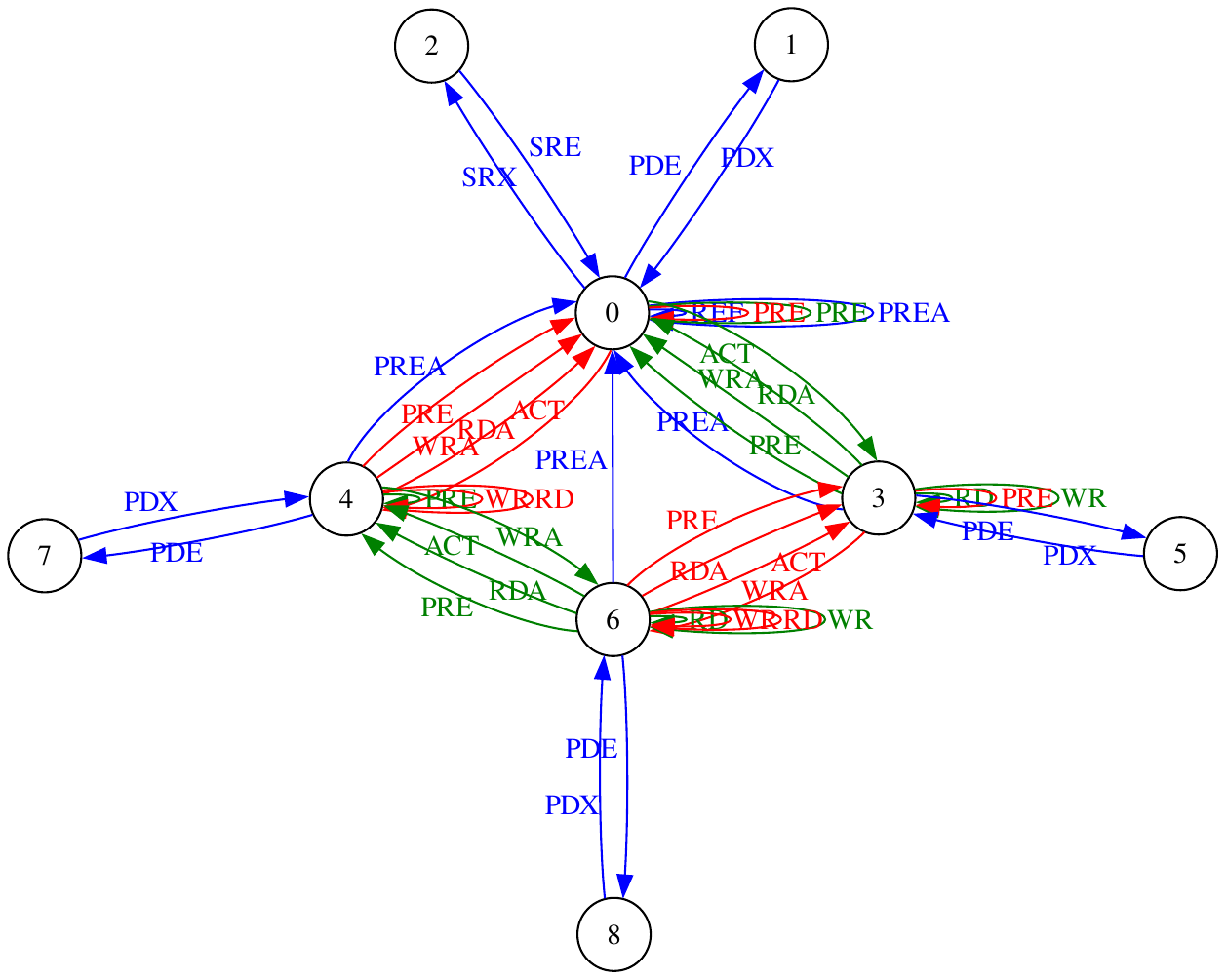}
    \caption{Traversal of the transitions and unrolling of the Petri net. Transitions are colored whether they affect \textcolor{OliveGreen}{Bank~0}, \textcolor{Red}{Bank~1} or the whole \textcolor{RoyalBlue}{Rank}.}
    \label{fig:traversal}
\end{figure}
In total, there are nine distinct states: the idle state (0); the two states in which only one bank is activated (3 and 4); and the state in which both banks are activated (6).
In state 6, both banks can issue \texttt{RD} and \texttt{WR} commands, which is the reason why two instances of these transitions are displayed in the graph.
The DRAM can go into power-down (1, 5, 7 or 8) from each of the former states.
Additionally, there is self-refresh (2) which can only be reached from the idle state.
In general, the number of distinct states can be written as $N = (2^B + 2^B + 1)^R = (2^{B+1} + 1)^R$, where $B$ is the number of banks and $R$ is the number of ranks.
As illustrated in Figure~\ref{fig:states}, this number quickly results in a state explosion when the number of banks or ranks increases.
\begin{figure}
\subfloat[R=1]{%
\begin{tikzpicture}
\begin{axis}[
    width = 0.5\columnwidth,
    height = 3cm,
    ybar,
    bar width=5pt,
    xlabel={Number of Banks},
    ylabel={Reachable States},
    xtick=data,
    ymin=1,
    ymode=log,
    xtick pos=left,
    ytick pos=left,
    symbolic x coords={1, 2, 4, 8, 16},
    enlarge x limits=0.15
]

\addplot coordinates {(1,5)(2,9)(4,33)(8,513)(16,131073)};

\end{axis}
\end{tikzpicture}
}
\subfloat[B=8]{%
\begin{tikzpicture}
\begin{axis}[
    width = 0.5\columnwidth,
    height = 3cm,
    ybar,
    bar width=5pt,
    xlabel={Number of Ranks},
    xtick=data,
    ymin=1,
    ymode=log,
    xtick pos=left,
    ytick pos=left,
    symbolic x coords={1, 2, 4, 8, 16},
    enlarge x limits=0.15
]

\addplot coordinates {(1,513)(2,263169)(4,69257922561)(8,4.79665983e21)};

\end{axis}
\end{tikzpicture}
}
\caption{Total number of reachable states for different quantities of banks and ranks.}
\label{fig:states}
\end{figure}
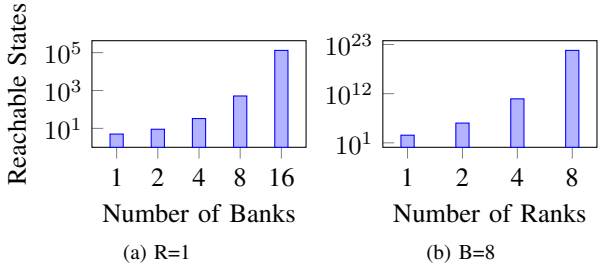
Assuming only one rank, $N(R=1,B=8)=513$ corresponds to DDR3 and $N(R=1,B=16)=131073$ corresponds to DDR4.

By examining the transition paths between states, this analysis also reveals an interesting metric, $k_{min}$, which represents the minimum number of transitions required to reach each possible state.
In our example, the longest command path leads to state 8, where both banks are first activated and then the DRAM is switched to power-down mode, resulting in $k_{min} = 3$.
In general, the longest path can be calculated as $k_{min} = (B + 1) \cdot R$ since all ranks must reach the active power-down state sequentially.

\subsection{Verification}
The executable Petri net model can also be used for verification.
For example, it can be used to verify the functional correctness of memory controllers with respect to the DRAM protocol.
Here, the \ac{dut} can be both an RTL model or a DRAM simulator in an \ac{uvm} setting with the executable model as a scoreboard.
The scoreboard replays the command traces fuzzily generated by the \ac{dut} and accurately reflects the DRAM’s internal state, flagging any illegal state transitions as soon as they occur.
Similarly, the model can be used to verify not only the controller but also DRAM logic:
Using the generated k-deep traces of all valid command sequences allows the DRAM logic to be fed command traces that achieve $100\%$ coverage of all scenarios for the given depth $k$ from a functional perspective.

In addition, as previously explained, the set of all possible command state transitions for a given trace depth of $k$ is a potential indicator for verifying the equivalence or similarity of two Petri nets.
Analyzing the $k_{min}$-deep traces of Petri nets across various DRAM standards highlights their differences and identifies where modifications to the RTL model are necessary.
Figure~\ref{fig:command_traces} shows the experimentally gained number of all possible command traces for a given depth $k$, as well as the algorithms runtime, for the exemplary DRAM.
\begin{figure}
\subfloat[Number of Valid Traces]{%
\begin{tikzpicture}
\begin{axis}[
    width = 0.5\columnwidth,
    height = 3cm,
    ybar,
    bar width=5pt,
    xlabel={$k$},
    ylabel style={align=center},
    ylabel={Count},
    xtick=data,
    xtick pos=left,
    ytick pos=left,
    ymin=1,
    ymode=log,
    enlarge x limits=0.15
]

\addplot coordinates {(1,8)(2,52)(3,368)(4,2664)(5,19624)(6,145926)(7,1091106)};

\end{axis}
\end{tikzpicture}
}
\subfloat[Algorithm Runtime]{
\begin{tikzpicture}
\begin{axis}[
    width = 0.5\columnwidth,
    height = 3cm,
    ybar,
    bar width=5pt,
    xlabel={$k$},
    ylabel style={align=center},
    ylabel={Runtime [s]},
    xtick=data,
    xtick pos=left,
    ytick pos=left,
    ymode=log,
    log origin=infty,
    enlarge x limits=0.15
]

\addplot coordinates {(1,105e-3)(2,109e-3)(3,138e-3)(4,315e-3)(5,1.655)(6,11.8)(7,90.4)};

\end{axis}
\end{tikzpicture}
}
\caption{Number of valid command traces for the depth $k$ and runtime of the exploration algorithm.}
\label{fig:command_traces}
\end{figure}
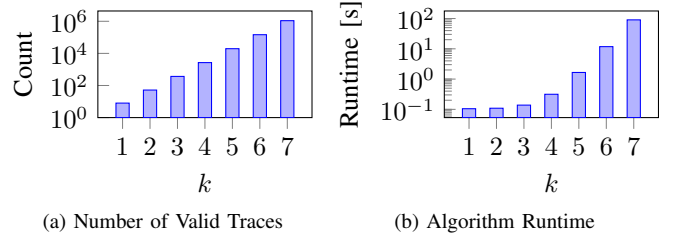
With $k_{min}=3$, there are $368$ possible command traces.
After that, the command sequence will only show cases that have already been explored.
Listing~\ref{lst:command_traces} presents an excerpt of generated command traces for $k=3$, showing a subset of all valid command sequences.
\begin{listing}
\footnotesize
\begin{lstlisting}
[PREA  (RA0)    ; PREA  (RA0)    ; PREA  (RA0)   ],
[PREA  (RA0)    ; PRE   (RA0BA1) ; SRE   (RA0)   ],
[ACT   (RA0BA0) ; ACT   (RA0BA1) ; WR    (RA0BA0)],
[SRE   (RA0)    ; SRX   (RA0)    ; PDE   (RA0)   ],
[ACT   (RA0BA0) ; PREA  (RA0)    ; ACT   (RA0BA0)],
[ACT   (RA0BA0) ; ACT   (RA0BA1) ; PDE   (RA0)   ],
                        ...
\end{lstlisting}
\caption{Sample command traces with $k=3$.}
\label{lst:command_traces}
\end{listing}

\subsection{Source Code Generation}
Since the Petri net model is built entirely using Python data structures, it is easy to iterate through the graph.
The timing constraints, as defined in Listing~\ref{lst:pyml_timing_constraint}, are used to dynamically add timing arcs to the Petri net while adhering to the specified component level.
Once all timing arcs are instantiated, the model is ready for code generation:
For example, we generate DRAMSys~\cite{stejun_22a} source code with templating libraries that automatically apply all defined command constraints based on the timing arcs in the Petri net.
With this method, we can introduce complex dependencies between commands in a flexible way and have them reflected in generated, correct-by-construction application code concerning protocol-complicance.
Similarly to~\cite{stesud_22}, we can also use the same concept to generate \ac{sva} properties.

\section{Conclusion}
\label{sec:conclusion}

In this paper, we presented a novel formal description method for DRAM protocols based on Python and timed Petri nets.
We demonstrated how this model can be made executable and, in doing so, constructed a basic memory controller that issues only legal command sequences.
Furthermore, we leveraged a set of algorithms for the Petri net model to conduct a series of property analyses.
We demonstrated how a set of all k-deep legal command sequences can be used to compare two Petri nets of different standards.
Additionally, we demonstrated how the Petri net model can generate source code for various applications while adhering to complex timing constraints.
Through this work, we aim to establish a standardized approach in expressing DRAM protocols using timed Petri nets.
We plan to release the Python data structures and DRAMPyML descriptions of several standards as open source.

In future work, it might be beneficial to wrap the flexible Python model again in a \ac{dsl} representation.
Consequently, the degree of freedom in writing code to represent a Petri net decreases, making it easier to visually inspect the source code and compare two Petri nets. Additionally, integrating timing into command traces will create a more expressive memory controller, hence a more robust way of proving or disproving timed Petri net similarities.
As a next step, we plan to model a large set of JEDEC DRAM standards using the new framework.

\bibliographystyle{IEEEtran}
\bibliography{ref}

\end{document}